# Monolithic superconducting emitter of tunable circularly polarized terahertz radiation

A. Elarabi[1,2,*], Y. Yoshioka[1], M. Tsujimoto[1,2], and I. Kakeya[1,†]

[1]Department of Electronic Science and Engineering, Kyoto University, Nishikyo, Kyoto 615-8510, Japan.
[2]Faculty of Pure and Applied Sciences, University of Tsukuba, 1-1-1 Ten-nodai, Tsukuba, Ibaraki 305-8573, Japan.

Abstract

We propose an approach to control the polarization of terahertz (THz) radiation from intrinsic Josephson-junction stacks in single crystalline high-temperature superconductor $Bi_2Sr_2CaCu_2O_8$. Monolithic control of the surface high-frequency current distributions in the truncated square mesa structure allows us to modulate the polarization of the emitted THz wave as a result of two orthogonal fundamental modes excited inside the mesa. Highly polarized circular terahertz waves with a degree of circular polarization of more than 99% can be generated using an electrically controlled method. The intuitive results obtained from the numerical simulation based on the conventional antenna theory are consistent with the observed emission characteristics.

---

[*]Corresponding author. asemelarabi@sk.kuee.kyoto-u.ac.jp
[†]Corresponding author.  kakeya@kuee.kyoto-u.ac.jp

## I. INTRODUCTION

The terahertz frequency range (0.3–10 THz) has attracted a considerable amount of interest in recent years. This is largely because of the numerous vibrational and rotational molecular absorption lines it contains which are used as marking regions in spectroscopy applications [1]. Ever since the demonstration of an intense THz radiation from a stack of intrinsic Josephson junctions (IJJs) made from $Bi_2Sr_2CaCu_2O_{8+\delta}$ (Bi-2212) [2], there has been an increased interest in the development of IJJ-based devices as highly promising emitters of coherent continuous-wave terahertz radiations [3–5]. Imaging, sensing, and spectroscopy applications require some control over the properties of radiations emitted from THz sources. The polarization of terahertz radiations is of special importance in vibrational circular dichroism spectroscopy (VCD) [6,7], high-speed telecommunications, and THz continuous–wave polarization imaging [8]. In laboratory setups, circular polarization (CP) can be realized through use of optical devices, such as quarter-wave plates; however, the use of compact monolithic sources has proven to be less costly and more suitable for applications that require portability. Circularly polarized THz emissions have recently been experimentally achieved by utilizing monolithic devices, such as quantum cascade lasers [9,10] and resonant-tunneling diodes [11]. Compared to these devices, IJJ-based sources can cover wider frequency ranges, which are unattainable by other THz sources [4]. Furthermore, the emission intensity of IJJ-based devices can also be thermally controlled, both internally [12,13] and externally [14,15]. By applying a direct current (DC) voltage ($V_b$) across the *c*-axis of IJJs stacked in a mesa form, a high frequency (HF) current is generated along with emission of sub-terahertz radiation at the Josephson frequency given by $f_J = (2eV_b/hN)$, where $h$, $e$, and $N$ denote the Planck constant, elementary charge, and number of contributing IJJs, respectively. Emission of high-intensity radiation is observed when the oscillating IJJs resonate in the excited cavity mode [2,3]. Extant studies have investigated the excited cavity modes in Bi-2212 mesa structures by observing the distribution of the far-field intensity as a function of the azimuth angle [16,17]. Additionally, the patch antenna model [18,19] has been found useful in describing the coupling between space and Josephson plasma waves inside the mesa structure [20–24]. A more direct approach to determine the excited cavity mode is to measure the polarization of the emitted radiation as a function of the mesa shape and bias conditions. These factors have, thus far, remained uninvestigated.

Waves emitted from rectangular IJJ-based sources have been known to be linearly polarized [2]. However, recent numerical studies [14,25,26] have proposed the possibility of realizing circularly polarized emissions from IJJ devices. This paper reports the first experimental demonstration of the generation of circularly polarized sub-THz waves from a monolithic source—a superconducting IJJ stack—wherein the ellipticity and evolution direction are tuned by mesa shape and bias conditions. The highest degree of circular polarization (DOCP) recorded in the study proposed herein is 99.7%, which is higher than that recorded by state-of-the-art quantum cascade lasers used for obtaining circular polarization [9,10,27]. As circularly polarized THz waves could be employed in a wide variety of potential applications, the result obtained in this study, along with recently reported packaging [28] and demonstration of THz torch [29], it is believed, will significantly influence and inspire the development of practical applications of IJJ-based THz emission devices [4,30].

## II. EXPERIMENTAL METHOD
### A. Samples design & fabrication

The mesa-shaped design is based on the truncated-edge square microstrip patch antenna [18,31–33]. Figures 1(a) and 1(b), respectively, depict a schematic of the IJJ-based device developed in this study and a microscopic image of sample #2. The mesa structure was fabricated from a Bi-2212 single crystal using photolithography and Ar-ion milling [13]. Circular polarization in the proposed design was realized through the excitation of two transverse orthogonal modes along the major and minor diagonals, as depicted in Figs. 1(c) and 1(d). On account of the small area of the truncated edge, the two above-mentioned transverse modes exhibit adjacent resonant frequencies. The mesa shape was designed with a perturbation to mix the two modes at the cavity resonant frequency (i.e., the arithmetic mean of the two adjacent resonant frequencies) with a phase difference of $\pi/2$ radians between them thereby forming a rotating surface current (refer to the animation in the supplementary data). According to the antenna theory, electric fields corresponding to the two above-mentioned eigen modes could be mathematically represented by $E_1 = (\sqrt{S}/w)(\sin kx - \sin ky)$ and $E_2 = (\sqrt{S}/w)(\sin kx + \sin ky)$ [31,32], where $w$ is the total length of the square side and $k = \pi/w$. The most significant design parameters that control the polarization state of the emitted radiation comprise the untruncated surface area $S$, the trimmed surface area $\Delta s$ (perturbation), and the feeding electrode position, as depicted in Fig. 1(d). The resonant frequencies of these two modes are given by $f_1 = f_{0r}$ and $f_2 = f_{0r}(1 + 2\Delta s/S)$ with $f_{0r} = c_0/2nw$ [18,32,34], where $c_0$ is the speed of light in vacuum and $n$ is the refractive index. The electric field distribution of the mixed mode can be represented by the linear combination of the two above-mentioned modes, as given below.

$$E_z = E_1 . e^{i2\pi f_1 t} + E_2 . e^{i(2\pi f_2 t + \delta)}, \tag{1}$$

where $\delta$ is the phase difference between the two modes. The above equation demonstrates that the first mode is antisymmetric with respect to the minor diagonal axis, and the second mode is antisymmetric with respect to the major diagonal axis, as depicted in Fig. 1(d). Therefore, the critical current density distribution $j_c$ must be antisymmetric with respect to both axes for the two modes to be excited at the same time. This implies that the feeding point must be on either the *x*- or *y*-axis for circular polarization to be achieved [35].

As oscillations in the proposed device are generated locally by Josephson junctions, the local rise in mesa temperature due to DC current injection acts as a perturbation to address the degeneracy of the two CP modes [36]. Antenna theory predicts that the position of an electrode, feeding HF from the source, with respect to the truncated edges determines the direction of rotation of the emissions [left-hand CP (LHCP) or right-hand CP (RHCP)] [18,37]; This implies that when the truncated edge is on the left side of the feeding point, RHCP emission is obtained and vice versa. Nonetheless, as mentioned earlier, the high-frequency feeding in the device under study is generated inside the mesa structure on account of DC current injection. This might altogether result in a rotational direction that may be different from the ones described above; this has been discussed later in this paper.

## B. Electromagnetic simulation

The initial design parameters were determined using the commercial full-wave three-dimensional finite-element electromagnetic simulation software—Ansys HFSS [26]. Similar simplified methods have previously been used to determine cavity resonance conditions [26,38], and a more advanced method, capable of solving sine–Gordon equations, was used to determine polarization properties [25]. A model similar to that reported in [26] was used, wherein the dielectric constant $\epsilon = 17.6$, and the feeding point is placed over the horizontal axis of the geometry. To account for the losses incurred owing to the skin effect, an impedance boundary condition was applied to the model [39], and the complex surface impedance was estimated by using, both, the Zimmermann method [26,39,40] and the *a–b* surface resistance correction [41] given by $R_s = \frac{1}{2}\mu_0^2 \sigma' \omega^2 \lambda^3$, where $\sigma' = 2/\omega\mu\delta_s^2$, $\delta_s$ is the skin depth, $\omega$ is the angular frequency, $\mu$ is the magnetic permeability, and $\lambda = 210$ nm is the penetration depth.

Figure 2(b) depicts a comparison of the axial ratios (ARs) determined through simulation of the proposed model and experimental results obtained from sample #3. The proposed model had parametric dimensions comparable to those of sample #3. Dimensions of the proposed model are $a_1$ = 13 μm and $a_2$ = 69 μm, where $a_1$ and $a_2$ represent the side length of the isosceles right-angled triangle-shaped truncated corner of the square microstrip and length of the remainder of the truncated square side, respectively, as shown in Fig. 1(d). The scattering parameter (S$_{11}$), as determined from the electromagnetic simulation, is shown on the right axis as a green line and demonstrates sharp minima at the two resonant frequencies $f_1$ and $f_2$.

## C. Measurement setup and polarization characterization

Samples were mounted on a Cu cold finger placed inside a He-flow cryostat with a transparent optical window, as depicted in Fig. 2(a). The radiation characteristics were detected by using a Si-bolometer and lock-in amplifier. The polarization was then measured by rotating a wire-grid polarizer in the beam path and recording the angle-dependent intensity. To characterize the polarization of the emissions, the polarization ellipse [10,42] was obtained by using the detected intensity as a function of the polarizer angle. The polarization state is best represented by the axial ratio for the entire emission range. The axial ratio, in dB, is defined as the ratio of the lengths of the major and minor axes (maximum and minimum intensities) of the polarization ellipse fitted to the polar plot of the detected intensity as a function of the polarizer angle; i.e., $\text{AR} = 20\,[\log(I_{\max}/I_{\min})]$. It is to be noted that polarization with an AR of less than 3 dB can be regarded as CP [37]. Frequency measurements were performed using a high-resolution lab-built Martin–Puplett FTIR interferometer [43], wherein the estimated emission frequency was obtained through linear interpolation of the measured data.

## III. RESULTS AND DISCUSSIONS

To discuss the effect of design parameters on the polarization state, we present the emission features of three samples with different $a_2/a_1$ values. The values of $a_1$ and $a_2$ and the thickness *t* of the mesa structure for the three samples are summarized in Table 1. *I–V* characteristics (IVC) and device emissions (pre-characterization) were first confirmed followed by polarization measurement through use of the above-mentioned setup [Fig. 2(a)].

Sample #1, along with all other samples, exhibits typical characteristics of slightly under-doped Bi-2212 mesas. IVC of sample #1 is depicted in Fig. 3(a) along with its detection intensity measured at the outermost IVC branch. In this study, all polarization measurements as well as IVC determination were performed at the bath temperature corresponding to the highest emission intensity ($T_b$ = 21 K). At about 13.4 mA, a sharp voltage dip is observed from 1.63 to 1.40 V, which could arguably be attributed to the formation of hotspots [12,44–47]. Emission is observed at a bias voltage $V_b = 1.6$ V and current $I_b = 10.4$ mA with two peaks—9.8 mA at 1.69 V and 8 mA at 1.89 V—as indicated by red arrows in Fig. 3(a). These peaks, at frequencies of 544.3 and 608.6 GHz, may correspond to the excitation of the two cavity modes [25]. The detected power (*P*), estimated in accordance with bolometer sensitivity, is found to be of the order of $P_{\max} \approx 176.5$ nW.

In the present study, the highest value of CP with AR = 0.2 was obtained under $I_b = 8$ mA and $V_b = 1.9$ V, as depicted in Fig. 3(b). At the point, DOCP was calculated to be 99.7%. This is the highest DOCP value recorded till date while using a monolithic THz source [10,9]. The AR increases accompanying considerable fluctuations as the emission diverge from the optimum bias point. Circularly polarized emission with AR < 3 dB is also observed in sample #2 as seen Fig. 3(c). The lowest AR recorded while using sample #3 was 4.5, which represents an elliptical polarization. Details of the obtained results are summarized in Table 1.

TABLE I. Design and experimental parameters as well as radiation properties of the three samples used in this study

| | Sample | #1 | #2 | #3 |
|---|---|---|---|---|
| Design parameters (μm) | $a_1$ | 16 | 20 | 13 |
| | $a_2$ | 70 | 76 | 69 |
| | w | 86 | 96 | 82 |
| | $a_2/a_1$ | 4.38 | 3.8 | 5.3 |
| | t | 2.25 | 1.9 | 2.4 |
| Temperatures (K) | $T_b$ | 21 | 22 | 40 |
| | $T_c$ | 84 | 83.3 | 82 |
| Radiation properties | $P_{max}$ (nW) | 176.5 | 23.5 | 123.5 |
| | $AR_{min}$ (dB) | 0.2 | 0.49 | 4.6 |
| | DOCP (%) | 99.7 | 99 | 77.8 |

The detection of the circularly and elliptically polarized waves is a clear indication of emission attributed to the flow of an in-plane superconducting current on the mesa surface. In a biased Josephson junction, an HF current flows across the barrier at a frequency determined by the AC Josephson relation. A standing wave of half a wavelength is consequently formed with antinodes located at both edges of a Josephson junction. Owing to the synchronization among the excited Josephson junctions in the stack, a net surface current oscillates according to the AC Josephson effect. This current gives rise to an oscillating magnetic field, which couples to surrounding space. In rectangular IJJ mesa devices, whose emissions have been intensively investigated by different groups, the synchronized transverse Josephson plasma wave that causes linearly polarized far-field detection is usually a two-dimensional (2D) standing wave polarized in the direction of z-axis and propagating along the x-axis [2,48]. In circularly polarized radiation, the propagation direction of the synchronized transverse Josephson plasma wave rotates at a frequency determined by the AC Josephson relation and resonance frequency relations of the two modes, as shown in Fig. 1(c), where $f_1\ and\ f_2$ in Eq. (1) correspond to Josephson oscillation frequencies.

By directly using the relation between the measured polarization-dependent intensity and the polarization ellipse [10,42], the polarization characteristics of sample #1 were determined. Figure 3(d) captures the polarization-dependent intensity at the minimum AR point in the polar plot. The detected intensity does not depend on *Θ*, although slight deformations in the two-fold symmetry may be detected. For samples #2 and #3, minimum AR values obtained in the same manner are 0.5 and 4.5, respectively. The variation in the value of minimum AR with respect to sample geometry is discussed in subsequent paragraphs.

As observed in Fig. 3(b), AR exhibits no significant dependence on bias-current and maintains a value less than 1 dB, with a fluctuation of about 0.5 dB, around the minimum AR condition (6.5–10.5 mA). At higher values of $I_b$ (> 10.5 mA), the value of AR is seen to rapidly increase. This is a common feature of samples #2 and #3. It may be concluded that when AR > 2 dB, its value is seen to rapidly increases with the deviation of $I_b$ from the optimum bias point. For sample #2, the Josephson frequency estimated by $V_b$ is plotted in Fig. 3 (c) along with the current evolution of AR. Any change in $V_b$ results in a corresponding change in the frequency of emission. This effect is believed to be caused by the detection of no signs of variation in the value of *N*, which usually appears as a discontinuity in measured $V_b$, during the experiments performed in this study. It was, therefore, considered that the estimated emission frequency varied within the range 0.46—0.50 THz with AR < 3 dB. Results obtained through numerical simulations indicate a steep dip in the minimum value of AR, as depicted in Fig. 4(b) [26]; on the other hand, however, experimental results indicate that AR does not demonstrate a strong dependence on frequency. This discrepancy in the results is believed to be caused by the trapezoidal shape of the mesa. The estimated geometrical resonant frequency exhibited by the geometry at the top of the stack was found to be 13% higher than that exhibited by the geometry at the bottom. This is supposed to be caused by an entrainment effect, wherein synchronization occurs between thousands of stacked IJJs at frequencies determined by the geometry of a part of the stack. Accordingly, the frequency range over which circular polarization is achieved is found to expand remarkably, and a sharp increase in AR is observed; this is in addition to the synchronization frequency range obtained through experiments.

In accordance with the antenna theory, the minimum value of AR (AR$_{min}$) in a truncated edge patch antenna can be achieved when $a_2/a_1 = \sqrt{2Q_0} - 1$ [31,32]. Hence, the two main factors that contribute to the attainment of CP are the ratio $a_2/a_1$ and the quality factor ($Q_0$). Results of the electromagnetic simulation indicate that the existence of an optimum value of $a_2/a_1$ that corresponds to the lowest value of AR$_{min}$. Figure 4(a) depicts the variation in AR$_{min}$, including the values calculated in accordance with the preconditions mentioned in section II (B) as well as those obtained experimentally in this study, as a function of $a_2/a_1$. The disagreement between optimum values of the ratio $a_2/a_1$ determined through simulation and experimental measurements may be attributed to the use of a simplified model. The value of $Q_0$, estimated from experimental results is 13 and that calculated through simulations is 36. The former value includes the ambiguity of dimensions on account of the trapezoidal shape of the mesa in addition to intrinsic material parameters, such as refractive index and surface impedance, which strongly influence the calculated value of $Q_0$.

Figure 5 depicts the emission spectra obtained for sample #3. Experimentally obtained AR as a function of the measured emission frequency, along with numerically calculated AR is plotted in Fig. 2 (b). The observed deviation from model calculation is believed to have been by reasons similar to the ones mentioned in the preceding paragraph. The measured frequency range (0.435–0.457 THz) closely agrees with the estimated modal frequency for geometries occupying the top of the mesa stack [49]. Also, similar to the case illustrated above, $V_b$ was found to be considerably higher than its value predicted using the emission frequency. This is believed to be due to excess voltage along the *ab*-plane accompanied by nucleation of hotspots [50]. as depicted in Fig. S1 in the supplementary material. Accordingly, the measurement of electric potential at the edges may result in higher values of the estimated Josephson frequency. Asymmetric electrode positions and emissions corresponding to the back-bending region of the IVC may also yield similar results as frequency estimation is also affected by Joule heating [49]. Frequency measurements performed for sample #3 and previously published experimental results [38] indicate that electromagnetic simulations can be used to identify cavity resonant frequencies. Figures 4(b) depicts the variation of the scattering parameter ($S_{11}$) (blue line) and AR (red line) with the frequency of emission for sample #1. On the basis of these results, the resonant frequencies for sample #1 are estimated as $f_1 = 402$ GHz and $f_2 = 434$ GHz.

Recently, Asai and Kawabata predicted the emission of circularly polarized radiation from a slightly rectangular IJJ mesa, wherein the rise in local temperature was achieved through laser irradiation [25,36]. Their studies claim that LHCP is achieved when a corner, located on the left of the major bisector and facing the center ("A"), is irradiated by laser radiation. By the same reasoning, RHCP would be achieved when the adjacent corner (either

"D" or "G") is heated. These results suggest that the in-plane superconducting current of the mesa surface, which causes a propagating electric field in the space, rotates in the direction defined by an arbitrary vector that starts from the point of irradiation and ends at the major bisector of the rectangular mesa (counter-clockwise and clockwise current rotations observed at points A and D, respectively). In the experiments conducted in this study, the current electrode provides heat to the mesa and ends up performing a function similar to that performed by laser irradiation in the above study. In the proposed study, sample #1 exhibits a similar trend in terms of shape and heating as the model used in the above-mentioned study. This similarity could be attributed to the fact that the heating portion is located on the right of the major (untruncated) diagonal of the mesa (facing the center of the mesa). Therefore, in sample #1 (truncated left edge), we consider that the in-plane current rotates counterclockwise; therefore, LHCP radiation is obtained under a bias of 7–10 mA. For the same reasons, the electric field generated using sample #2 [Fig. 1(b)] is left-handed and that generated using sample #3 (truncated right edge) is right-handed. Thus, the direction of polarization rotation predicted for truncated-edge square mesa is opposite to that estimated by antenna theory for similarly shaped antenna fed at the same position.

In order to better understand the progression of polarization over the entire bias current range, plots in Figs. 6(a and b) depict the observed current-dependent intensity distribution in the form of a false-colored 2D contour plot for samples #1 and #3, respectively. As can be realized in Fig. 6(a), the maximum emission intensity is achieved in the minimum AR (best CP) at current range of 8.3–8.5 mA; however, a rather low intensity is observed at the other low AR point (7 mA). In sample #1, the maxima of intensity correspond to the minimum values of AR (< 1 dB). It was also observed that the angle of the major axis, indicated by an asterisk, is independent of $I_b$. Considering that the observed emission is a superposition of the two circular polarized modes (RHCP and LHCP), the AR evolution could be attributed to the variation in amplitudes of the two circularly polarized modes with a constant phase difference between them. Some minor fluctuations in the measured angle of the major axis angle were noticed in sample #1, which correspond to a small change in magnitude of the transverse mode owing to the imperfect shape of the edges. This fluctuation was less noticeable in sample #3 [Fig. 6(b)]. The author intends to investigate this issue in detail in a future endeavor.

## IV. CONCLUSIONS

This paper presents the design, fabrication, and experimental demonstration of a technique to generate circularly polarized terahertz radiations that comprise low axial ratios and maintain a high output intensity through use of an intrinsic Josephson junction oscillator. Circular polarization is achieved by using a simple truncated-edge square mesa structure. It verifies the simplicity and applicability of the antenna theory alongside existing electromagnetic simulation methods in achieving the desirable polarization. Furthermore, because the proposed method is based on the excitation of IJJs within a mesa cavity, it avoids issues with the insertion loss associated with external polarimetric modulators, which may be substantial in the terahertz frequency range. The polarization characteristics have been discussed at length in terms of both the antenna theory as well as electromagnetic simulations. These results pave the way for the development of enhanced circularly polarized sources of terahertz radiation for use in practical applications, such as mobile communications and circular dichroism spectroscopy.

**ACKNOWLEDGMENTS**

The authors acknowledge the technical support from F. Tubbal. The Bi-2212 single crystals were provided by Y. Nakagawa and Y. Nomura. This work has been supported by KAKENHI No. 26286006 and Murata Science Foundation.

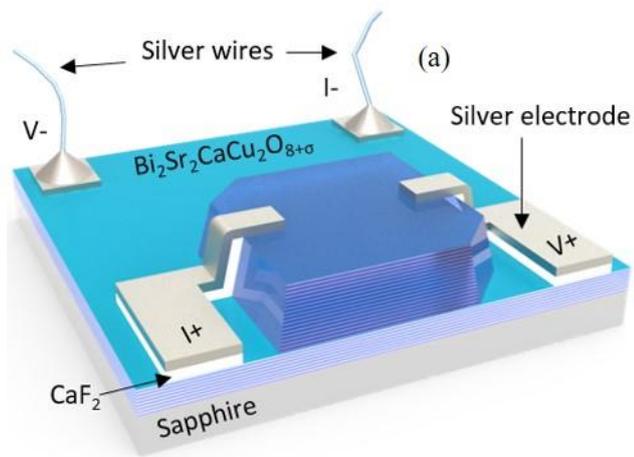
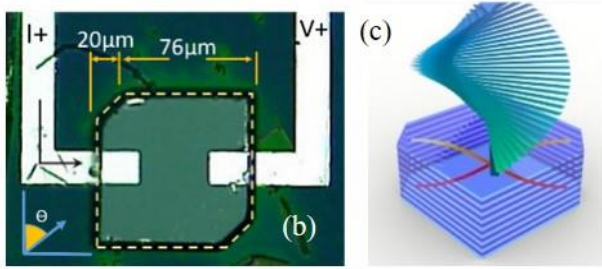
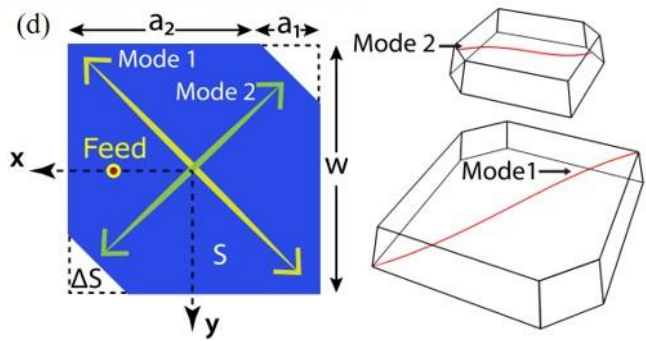

FIG. 1. (a) Schematic of Bi-2212 terahertz emitter. (b) Optical microscopic image of sample #2. (c) Electric-field distribution and circularly polarized radiation propagation and direction. (d) Excitation of the two orthogonal transverse modes along the major and minor diagonals in the x–y plane.

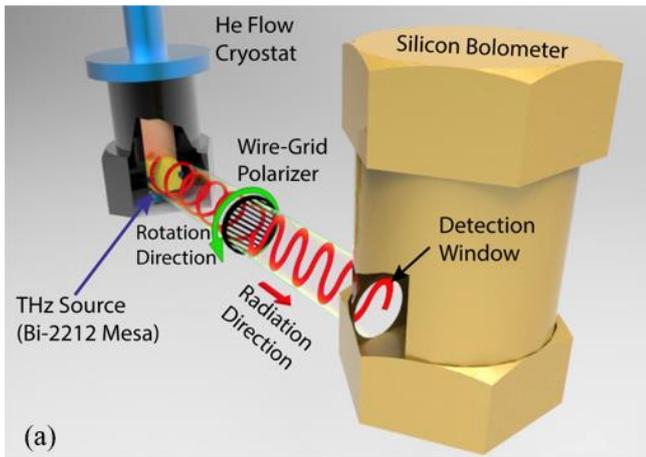
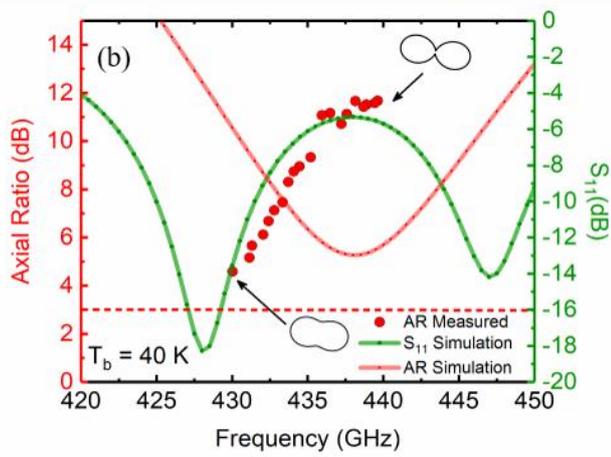

FIG. 2. (a) Schematic of the experimental setup for polarization characterization. (b) Axial ratio comparison between sample #3 and the results obtained from electromagnetic simulation.

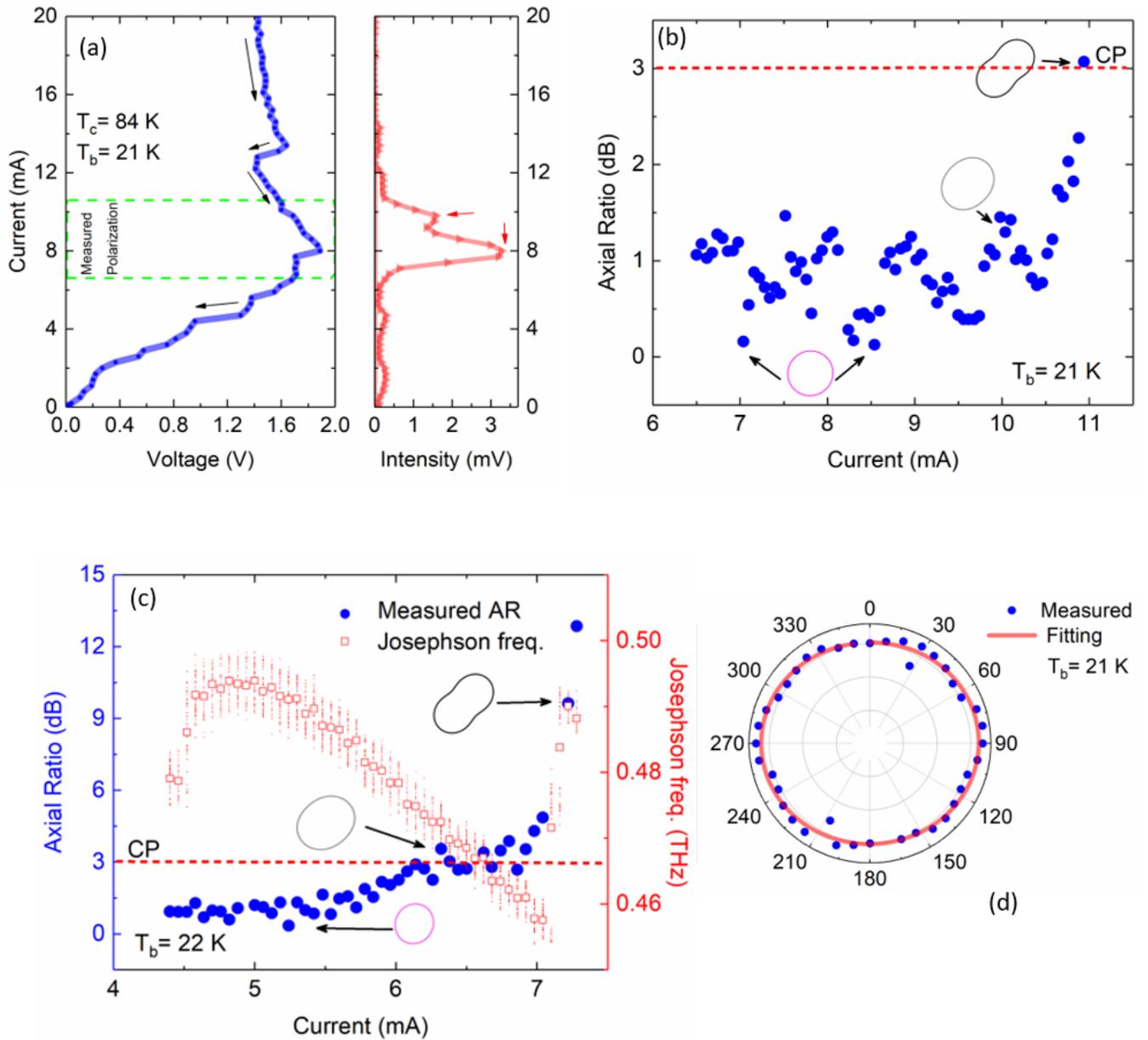

FIG. 3 (a) Return branch of the IVC (left) of sample #1 at $T_b$ = 21 K along with the detected emission intensity as a function of bias current (right); AR evolution (blue solid symbols) as a function of the bias current for (b) sample #1; and (c) sample #2; in (c), the red open symbols represent Josephson frequencies estimated from $V_b$ and small dots represent $V_b$ fluctuations during measurements; (d) Polar plot of the detected emission intensity as a function of the polarizer angle $\theta$. The data are extracted from Fig. 3(b) at the minimum AR value. The solid lines represent a sinusoidal fitting.

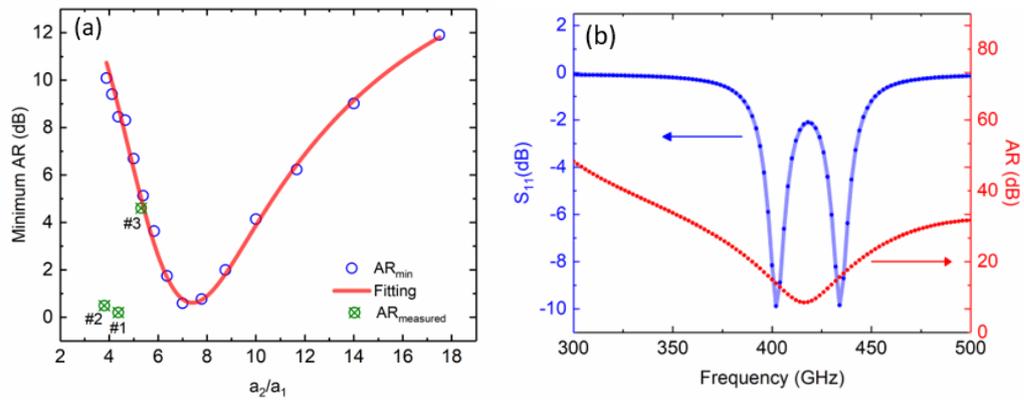

FIG. 4. (a) Minimum values AR vs. dimensional parameter demonstrating the optimum $a_2/a_1$ value for lowest simulated (AR$_{min}$) and compared to that corresponding the measured AR$_{measured}$. The red solid line is a polynomial fit as a guide for the eyes. (b) Electromagnetic simulation of AR and $S_{11}$ using the actual dimensions of sample #1.

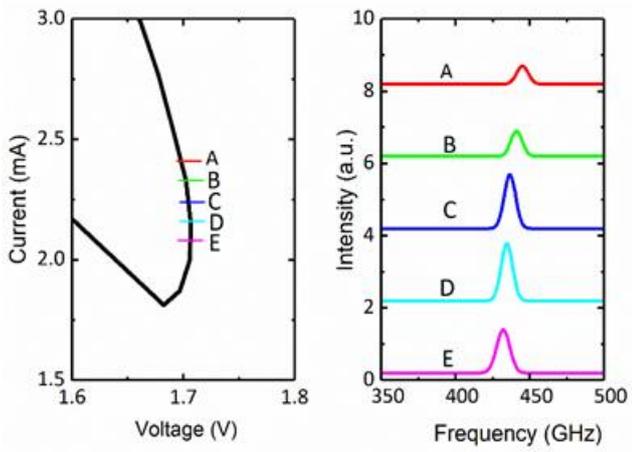

FIG. 5 Part of IVC (left) and emission spectra (right) of sample #3. Symbols marked A–E in the IVC represent bias conditions for respective spectra. Spectrum curves are vertically shifted to enable ease of visibility.

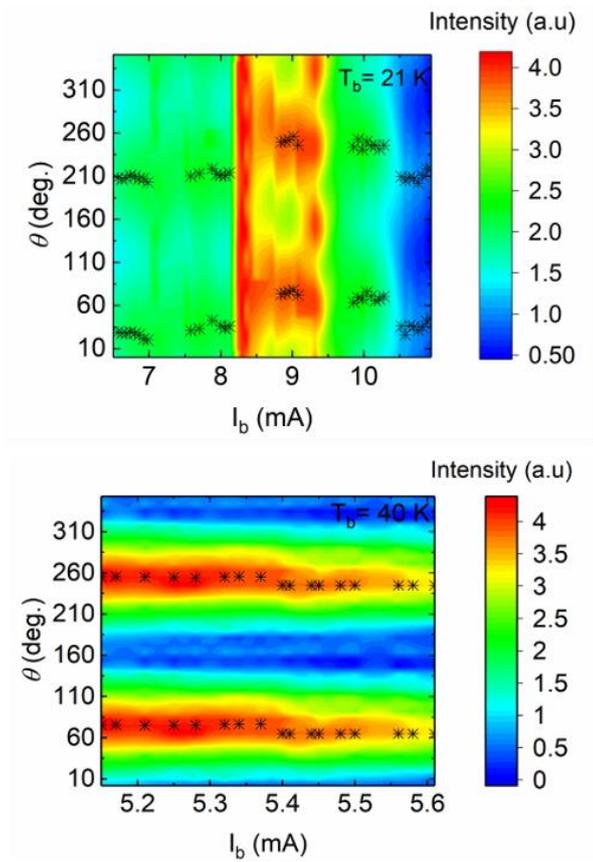

FIG. 6. Measured current- and angular-dependent intensities for the entire emission range of samples (a) #1 and (b) #3. The major axis angles corresponding to higher values of AR (> 1 dB) are represented by asterisks.